\documentstyle[aps,prb,amssymb,amsmath,graphicx]{revtex}
\begin{document}
\draft
\title{Fluctuating diamagnetism in underdoped high temperature superconductors}
\author{Alain Sewer$^{1,2}$ and Hans Beck$^{2}$}
\address{$^{1}$Institut Romand de Recherche Num\'erique en Physique des 
Mat\'eriaux (IRRMA), EPFL,\\ 1015 Lausanne, Switzerland\\
$^{2}$Institut de Physique, Universit\'e de Neuch\^atel, 2000 Neuch\^atel,
Switzerland}
\date{\today}
\maketitle
\begin{abstract}
The fluctuation induced diamagnetism of underdoped high temperature
superconductors is studied in the framework of the Lawrence-Doniach
model. By taking into account the fluctuations of the phase of the 
order parameter only, the latter reduces to a layered XY-model
describing a liquid of vortices which can be either thermally excited 
or induced by the external magnetic field. The diamagnetic response
is given by a current-current correlation function which is
evaluated using the Coulomb gas analogy. Our results are then
applied to recent measurements of fluctuation diamagnetism in 
underdoped YBCO. They allow to understand both the observed anomalous 
temperature dependence of the zero-field susceptibility and the
two distinct regimes appearing in the magnetic field dependence 
of the magnetization.  
\end{abstract}
\pacs{PACS: 74.20.De, 74.25.Ha, 74.40.+k}


\section{Introduction}
Owing to their short coherence length, high temperature superconductors 
show marked deviations from the mean-field behaviour that describes
rather well the behaviour of conventional superconductors. The
underdoped regime of the various cuprates is particularly interesting,
given their pronounced anisotropy and the low density of charge
carriers. In this region of the phase diagram, fluctuations are strongly
enhanced and are manifest already well above the critical temperature 
$T_c$. This fluctuation regime gives rise to various unusual 
phenomena,\cite{recrev} such as an anomalous temperature dependence of 
the Knight shift, NMR relaxation rate and electrical conductivity, an
anomalous frequency dependence of infrared conductivity, as well as the
formation of a pseudogap in the electronic density of states. Close to
$T_c$ it also allows to see true critical behaviour in quantities such
as the specific heat \cite{junod} or thermal expansion.\cite{pasler}
\\
In the present work, we specifically address the temperature and field
dependence of the diamagnetic susceptibility $\chi$ of strongly 
anisotropic high temperature superconductors above $T_c$. The influence 
of fluctuations on $\chi$ has been studied long ago in the framework of 
a Landau-Ginzburg model, taking into account Gaussian fluctuations 
of the order parameter above the critical temperature.\cite{schmid}
More refined calculations based on the Lawrence-Doniach
model have taken into account the lattice structure.\cite{baraduc} 
Recent measurements \cite{carretta} have shown that the Gaussian 
approximation can indeed well describe the diamagnetic fluctuations
in optimally doped YBCO, whereas underdoped specimens of the same 
compound, such as YBa$_2$Cu$_3$O$_{6.67}$, show dramatic deviations 
from this behaviour. The fluctuation region where the zero-field
orbital susceptibility shows appreciable values extends over a much 
larger temperature range than in the optimally doped system. Moreover, 
the field dependence of the magnetization shows a much more pronounced 
crossover between low and high fields than what would have been 
expected from Gaussian fluctuations.
\\
We base our calculations on an anisotropic Lawrence-Doniach (LD)
functional, involving the superconducting order parameter field 
$\Delta$ in the presence of a vector potential that describes a 
homogeneous magnetic field perpendicular to the lattice planes. 
Rather than considering Gaussian fluctuations, we assume a 
``precursor regime'' in which the amplitude of $\Delta$ has 
already acquired a non-zero average value whereas its phase is
subject to strong fluctuations inhibiting long range superconducting 
order. In this context the LD-functional reduces to an anisotropic 
layered XY-model. The relevant thermal excitations of such a 
system are the phase field singularities which manifest themselves
as vortices and antivortices in 2D and as vortex loops in 3D. 
The applied magnetic field also acts on the phases by inducing
vortex lines crossing the sample from one end to the other. The
diamagnetic susceptibility, expressed in the usual way by a
current-current correlation function, is then related to the 
positional correlation function of the vortex line elements, the
static structure factor $S({\mathbf q})$. We model 
$S({\mathbf q})$ in a simple way by using various known results 
obtained either by analytic considerations based on the Coulomb gas 
analogy or by Monte Carlo (MC) simulations of the anisotropic 3D 
XY-model. Then we arrive at explicit expressions for the zero-field 
susceptibility $\chi(T)$ and the temperature and field dependent 
magnetization $M(T,B)$. They contain several material dependent 
parameters which are estimated by comparing with the experimental
data on underdoped YBaCuO from Ref.6.
\\
This procedure unravels three main features. First, the density of 
the vortex line elements contributing to the diamagnetic response
(those which are oriented in $z$-direction, i.e., parallel to the
applied field) is thermally activated with a value of the
activation energy that is compatible with what is found in the
above mentioned MC simulations. Second, in the temperature range
covered by the experiments, the value of the anisotropy shows that
the positions of the vortex line elements fluctuate strongly from
one layer to the other. This points to a rather weak effective
coupling between layers, which is compatible with the fact of being
above the vortex melting, respectively the ``vortex decoupling''
line. Finally the rather sharp crossover of the magnetization
$M(T,B)$ as a function of $B$ points to a subtle interplay between
thermally excited vortex loops and field induced vortex lines.
\\
These observations allow for the following conclusions: 
\textit{(i)} The experiments in Ref.6 can be much better accounted 
for by phase excitations than by Gaussian fluctuations of the 
pairing field. This is particularly manifest in the activated 
$T$-dependence of the susceptibility, but also in the existence 
of two field regimes with quite different behaviours of the 
magnetization. In this context we have, however, to admit that the 
very sharp crossover between these two regimes observed in the
experiment may also be due to sample inhomogeneities as it was 
suggested in the experimental papers.\cite{carretta} \textit{(ii)} 
Except for the data taken close to the zero-field critical
temperature, the diamagnetic response of the underdoped compound
presented in Ref.6 seems to be ``precritical'', in the sense
that the relevant lengths in the lattice planes as well as in the 
perpendicular direction do not show any true critical (i.e., singular) 
behaviour (which would be supposed to belong to the 3D XY universality 
class \cite{pasler}).
\\
In section II we develop the theoretical formalism that allows to express
$\chi(T)$ and $M(T,B)$ in terms of the vortex line structure factor, and 
in section III we compare our theoretical results with the data presented 
in Ref.6 for underdoped YBCO, thereby extracting the free parameters from
the experimental curves. A summary is presented in section IV.

\section{Diamagnetic response}

\begin{figure}
\includegraphics[width=13cm]{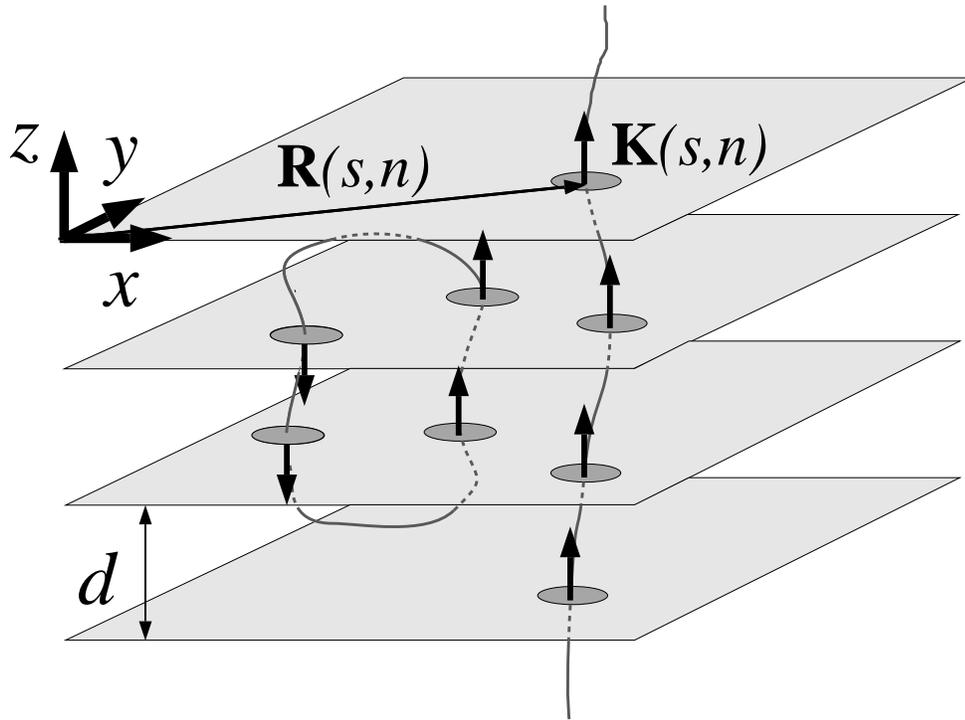}
\caption{Schematic representation of a 3D thermally excited vortex
loop (left) and of a field-induced vortex line (right). Since in the
Lawrence-Doniach approach the phase field is defined only for discrete 
values of the $z$-coordinate ($z=nd$), one often refers to these structures 
as ``stacks of pancake vortices''. The latter are represented by the grey 
ellipses whereas the lines linking them are then just guides to the eye.}
\label{fig1}
\end{figure}

We discuss the orbital magnetic response of an underdoped superconductor
in the London approximation to the Lawrence-Doniach (LD) model, i.e., 
in the framework of an 3D anisotropic layered XY-model in which the phase 
$\theta$ of the superconducting order parameter $\Delta$
is coupled to the vector potential ${\mathbf A}_{\parallel}$ describing 
a homogeneous magnetic field $\mathbf B$ perpendicular to the lattice
planes (we restrict ourselves to temperatures $T>T_c$ where the Meissner
effect is absent, identifying thus the external and the effective internal 
vector potential):
\begin{equation}
{\mathcal H}[\theta]=\frac{1}{2a^2d}\sum_n\int {d}^2\!r\, 
\{J_{\parallel}a^2 [{\boldsymbol\nabla}_{\!\parallel}\theta_n
-\frac{2\pi}{\Phi_0}{\mathbf A}_{\parallel}]^2
+J_{\!\perp} [1-\cos(\theta_n-\theta_{n+1})]\}.
\label{XYaction}
\end{equation}
Here $a$ is the lattice constant in the planes whereas $d$ denotes 
the distance between two layers. $J_{\parallel}$ and $J_{\perp}$ are 
the respective phase couplings. Their ratio
\begin{equation}
\gamma^2=\frac{J_{\parallel}}{J_{\!\perp}}>1
\end{equation}
determines the anisotropy of the system. The XY-Hamiltonian can be
obtained starting from a LD functional for the complex superconducting 
pairing field $\Delta$ by keeping the amplitude of the latter constant
(London approximation). This is a current strategy \cite{EK} based on the 
assumption that the various precursor phenomena, observed in underdoped 
cuprates between $T_c$ and some higher temperature $T^*$ and mentioned in 
the introduction, are essentially due to fluctuations of the phase of 
$\Delta$ whereas its amplitude $|\Delta|$ maintains a finite mean value. 
Moreover, these materials exhibit a layered structure which is 
specifically taken into account by the Lawrence-Doniach approach. 
Expression (\ref{XYaction}) is a partial continuum version of the 
discrete 3D XY-model, the Josephson coupling  
$1\!-\!\cos(\theta_i\!-\!\theta_j)$ between neighbouring lattice sites 
in a given layer having been replaced by the phase gradient. 
\\
The orbital magnetic response $\Lambda$ in a finite external 
field is obtained by adding a small perturbation to the applied 
vector potential
\begin{equation}
{\mathbf A}_{\parallel}\rightarrow{\mathbf A}_{\parallel}
+\delta{\mathbf A}_{\parallel}
\end{equation}
and by calculating the second derivative of the free energy with respect
to the perturbing field $\delta{\mathbf A}_{\parallel}$. The 
magnetic susceptibility $\chi(T,B)$ is then given by
\begin{equation}
\chi=\lim_{q\rightarrow 0}\frac{\Lambda({\mathbf q})}{q^2}
\label{limit}
\end{equation}
with
\begin{equation}
\Lambda({\mathbf q})=
\frac{J_{\parallel}}{d}\,\big(\frac{2\pi}{\Phi_0}\big)^2\,
[\frac{J_{\parallel}}{k_{\!B}T}\,C({\mathbf q})-1].
\label{ddJdAdA}
\end{equation}
The second term of (\ref{ddJdAdA}) is the diamagnetic 
response, whereas the first term involves the current-current 
correlation function
\begin{eqnarray}
C({\mathbf q})&=&\frac{1}{L^2}\sum_{n,n'}\int 
{d}^2\!\rho\,{d}^2\!\rho'\, 
e^{i{\mathbf q}\cdot({\mathbf r}-{\mathbf r}')}
\langle j_x({\mathbf r})j_x({\mathbf r'}) \rangle
\label{diamresp}
\\
 j_x({\mathbf r})&=&\nabla_{\!\!x}\theta_n({\boldsymbol \rho})
-\frac{2\pi}{\Phi_0} A_{\parallel,x}({\mathbf r}),
\label{current}
\end{eqnarray}
Here the coordinate $\mathbf r$ means $({\boldsymbol\rho},nd)$ and 
the sample volume is $\Omega=L^2Nd$ where $N$ is the number of layers. 
In order to have a finite $\chi$, the limit $C(q\rightarrow 0)$ has 
to cancel the diamagnetic term. It will be shown below that this is 
indeed the case in our approach. We have chosen the gauge in which
\begin{equation}
{\mathbf A}_{\parallel}({\mathbf r})=\left(-yB,0,0\right).
\end{equation}
Thus the relevant wave vector $\mathbf q$ has only a 
$y$-component denoted simply by $q$.
\\
In order to study the thermodynamic properties of the 3D anisotropic 
XY-model, we first recall that the Berenzinskii-Kosterlitz-Thouless 
transition occurring in the strictly 2D case is best described in terms
of vortex and antivortex excitations. Although the 3D XY-system shows a 
``normal'' second order transition, even when it is anisotropic, 
it has been shown that it is also possible to understand the critical 
behaviour of such a system in terms of vortex excitations, which 
- for topological reasons - now have to form closed loops or 
continuous lines crossing the whole system.\cite{chatto} The loops
are the 3D extension of the planar vortex and antivortex structure 
whereas the lines arise from the presence of an external magnetic 
flux penetrating into the sample in the same way as in a type II 
superconductor below $T_c$. In the following we will use this vortex
picture of the 3D XY-model in order to calculate the phase correlation
function $C({\mathbf q})$ that determines the diamagnetic response
according to Eq.(\ref{diamresp}). 
\\
Since vortex lines are either closed (forming a loop) or extend 
continuously through the whole sample, we can characterize their 
structure by labelling each line by an index $s$ and by giving its 
position ${\mathbf R}(s,n)$ in a given layer $n$ which corresponds
to the centre of the corresponding ``pancake vortex''
(see Fig.\ref{fig1}). The $x$-component of the phase gradient created
by all the vortex lines is then given by the same expression used in 
magnetostatics in order to calculate the magnetic field of a system 
of current loops and lines:
\begin{equation}
\nabla_{\!\!x}\theta_n ({\boldsymbol \rho})=d\sum_{\alpha,\beta,s,n'} 
\varepsilon_{x\alpha\beta} \int {d}^2\!\rho'\,
\frac{(r_{\alpha}-r_{\alpha}')\,K_{\beta}(s,{\mathbf r'})}
{|{\mathbf r}-{\mathbf r'}|^3}.
\label{biosavard}
\end{equation}
Here $\varepsilon_{\alpha\beta\gamma}$ is the fully antisymmetric 
tensor of rank 3 and the vector field 
${\mathbf K}(s,{\mathbf r}')$ is given by the line element
``tangential'' to the vortex line number $s$ at point 
${\mathbf r}' = ({\boldsymbol\rho}',n'd)$:
\begin{equation}
{\mathbf K}(s,{\mathbf r}')=t(s,n')\,\delta {\boldsymbol (}{\boldsymbol\rho}'-
{\mathbf R}(s,n'){\boldsymbol )}\,\hat{\mathbf z},
\label{ka}
\end{equation}
$\hat{\mathbf z}$ being the unit vector in the $z$-direction and 
the sum in (\ref{biosavard}) thus runs over all vortex line elements
$s$ and layers $n$. In order 
to make connection with the (more simple!) 2D case, it is useful to
attribute a topological number $t(s,n)=\pm 1$ to each vertical vortex line 
element. Its sign is chosen such that the product 
$t(s,n)\,\hat{\mathbf z}$ 
gives the oriented ``tangential'' vector of the vortex line at that 
point [i.e., $t=+1$ ($-1$), when the line moves upward (downward) with 
respect to the lattice plane $n$]. The current correlator (\ref{diamresp})
is then given by two contributions:
\begin{equation}
C({\mathbf q})=\frac{4\pi^2}{L^2 N q^2}\,
[S({\mathbf q})-
\big(\frac{L^2 N B}{\Phi_0}\big)^2\,\delta_{{\mathbf q},0}
].
\label{curcurcor}
\end{equation}
The first term, stemming from the phase gradient in the current density
(\ref{current}), represents the structure factor of the vortex line 
elements oriented in $z$-direction, given by
\begin{equation}
S({\mathbf q})=\sum_{s,n,s',n'}t(s,n)\,t(s',n')\,\langle
e^{i{\mathbf q}\cdot [{\mathbf R}(s,n)
-{\mathbf R}(s',n')]}\rangle.
\label{strucfac}
\end{equation}
The other term in (\ref{curcurcor}) comes from the second 
(diamagnetic) contribution to the current (\ref{current}) and 
represents the total flux going through the system due to the 
applied field (cross correlations between $\nabla_{\!\!x}\theta$ 
and $A_{\parallel,x}$ are supposed to vanish due to the disordered 
structure of the vortex system above the melting temperature). In 
the Appendix we show that the singular zero wave vector value of
the correlation function $S({\mathbf q})$ compensates the
second contribution to (\ref{curcurcor}) so that the limit
$C(q\rightarrow 0)$ is well behaved. The main problem left is thus
to find a suitable form for the regular part of the correlation
function $S({\mathbf q})$.
\\
First, we recall that we are interested in the fluctuation induced 
diamagnetism above the zero-field transition temperature $T_c$, a range 
in which recent experiments have been carried out.\cite{carretta} For a 
finite field one has thus to deal with a region in the $T$-$B$-phase diagram 
laying beyond the different possible transition lines which are 
discussed in the literature, such as the vortex melting line and 
the vortex decoupling line.\cite{blatter,tesanovic} Our structure factor 
$S({\mathbf q})$ thus has to describe a disordered vortex-liquid system 
in which strongly ``wrinkled'' vortex lines and loops
\cite{chatto2,tesanovic} go through the sample. We use the following
reasoning for obtaining an approximate form of the regular contribution
to $S({\mathbf q})$:
\\ \\
\textit{(i)} In the extreme limit of totally decoupled layers, each plane 
would have to be described by a 2D XY-model above 
the critical temperature, which is frustrated when an external magnetic 
field is applied. Its vortex structure would be the one of a neutral Coulomb
gas, more precisely of a mixture of a neutral two-component Coulomb gas 
(given by an equal number of thermal vortices and antivortices) and of a 
one-component gas (the field induced vortices) in a neutralizing background 
(given by the external flux). In the purely 2D case, the non-trivial part
of Eq.(\ref{strucfac}) thus reduces to 
\begin{equation}
\begin{split}
S_{2D}({\mathbf q})&=
N \sum_{s,s'} t(s,0)\,t(s',0)\,\langle
e^{i{\mathbf q}\cdot[{\mathbf R}(s,0)-{\mathbf R}(s',0)]}\rangle\\
&= N L^2 n_V S_C({\mathbf q}).
\end{split}
\end{equation}
Here we represented the positional structure factor involving a sum over
all vortex and antivortex positions by the Coulomb gas structure factor 
$S_C({\mathbf q})$. We use the following approximate form \cite{march}
\begin{equation}
S_C({\mathbf q})=\frac{q^2}{q^2+2\pi n_V q_V^2/k_{\! B}T}
\label{coulomb}
\end{equation}
where $n_V$  is the areal vortex density and the ``charge'' $q_V$ of each 
vortex is related to the in-plane phase coupling $J_{\parallel}$ by
\cite{cote}
\begin{equation}
q_V^2=2\pi J_{\parallel}.
\label{charge}
\end{equation}
The expression (\ref{coulomb}) yields the correct limiting behaviour
of $S_C({\mathbf q})$ for $q\rightarrow 0$ and should be valid
for temperatures not too close to $T_c$. Inserting expression 
(\ref{coulomb}) into Eq.(\ref{curcurcor}) gives a similar form of the
current-current correlation function $C(\mathbf{q})$ as used by Kwon and
Dorsey.\cite{kwon}
\\ \\
\textit{(ii)} For a strong anisotropy and for the considered temperature 
range the
effective interlayer coupling will be very weak. For evaluating expression
(\ref{strucfac}), we split the double sum over $n$ and $n'$ 
into two parts. In the first part, we take $n$ equal to $n'$ which yields
a 2D problem analogous to the one treated above. The corresponding 
contribution to (\ref{strucfac}) is therefore similar to the 
structure factor of point vortices in a single XY-plane:
\begin{equation}
S_1({\mathbf q})= N L^2 n_V S_C({\mathbf q})
\label{layer}
\end{equation}
where $n_V$ is now the areal density of \emph{vertical} vortex line
elements. The second part $S_2({\mathbf q})$ of the 
double sum in (\ref{strucfac}) contains $n$ and $n'$ involving
different layers. Here we use the fact that in the considered 
temperature range, the vortex loops and also the field induced 
lines have a very irregular shape. Indeed, the region in 
the $T$-$B$-plane we are interested in lies above any lattice layer 
decoupling line.\cite{blatter} This fact is formulated in 
Ref.7, by indicating a $T$-domain
\begin{equation}
\frac{|T-T_c|}{T_c} > \gamma^{1/\nu}
\end{equation}
with an XY-exponent $\nu\approx 0.6$, for which the lattice planes are 
practically decoupled. Therefore only those ones which are close enough
to another (in a sense to be specified below) are correlated and will 
contribute to $S_2({\mathbf q})$. We first expand formally in expression 
(\ref{strucfac}) the position ${\mathbf R}(s',n')$ of a the vortex 
line $s'$ in layer $n'$ with respect to its value in layer $n$. This 
gives 
\begin{equation}
S_2({\mathbf q})=\sum_{s,s',n,n'\neq n} t(s,n)\,t(s',n')\,\langle
e^{i{\mathbf q}\cdot[{\mathbf R}(s,n)-{\mathbf R}(s',n)
]}\,e^{-i{\mathbf q}\cdot{\mathbf u}(s',n'-n)}\rangle.
\label{t2q}
\end{equation}
Then we split the average bracket of (\ref{t2q}) and factor out from the
sum over $n'$ the first exponential which pertains to a given layer $n$ as
well as the topological numbers $t(s',n')$ which are equal to $t(s',n)$
for neighbouring layers $n$ and $n'$. This yields the previous result
(\ref{layer}). We are then left with correlations between the positions
$\mathbf R$ of a single vortex line $s'$ in layer $n'$ and in layer
$n\neq n'$. Assuming that such correlations are the same for all vortex
lines $s'$ and extend only over a distance $\xi_3=n_3d$, we obtain
\begin{equation}
S_2({\mathbf q})=2\,S_1({\mathbf q})
\,\sum_{n=1}^{n_3}\,
\langle e^{-i{\mathbf q}\cdot{\mathbf u}(n)}\rangle
\equiv 2\,S_1({\mathbf q})\,X({\mathbf q}).
\label{others}
\end{equation}
Here ${\mathbf u}(n)$ is the deviation of a given line or loop from a 
straight line along the $z$-direction. The factor 2 comes
from the fact that the sums over $n<n'$ and $n>n'$
have been reduced to one such sum in (\ref{others}). Assuming that 
the vortex lines behave like harmonic strings with an effective 
stiffness given approximately by 
$\lambda\approx\frac{J_{\!\perp}}{d}=\frac{J_{\parallel}}{d\gamma^2}$,
one finds
\begin{equation}
\begin{split}
X({\mathbf q})&\approx\sum_{n=1}^{n_3}
\exp{\boldsymbol (}-\frac{1}{2}\,q^2\langle {\mathbf u}^2(n)
\rangle{\boldsymbol )}\\
&=\frac{1-\exp(-{k_{\!B}Tn_3 d\,q^2}/{4 \lambda})}
{\exp(-{k_{\!B}T d\,q^2}/{4\lambda})-1}.
\end{split}
\label{sergeo}
\end{equation}
Here we used the relation
$\langle {\mathbf u}^2(n)\rangle 
=\frac{1}{2}\,\frac{k_{\!B}T}{\lambda}\,n d$ applying for harmonic 
deformations and yielding a geometric series. The final result for the 
regular part of $C({\mathbf q})$ in (\ref{curcurcor}) then takes 
the form
\begin{equation}
C({\mathbf q})=\frac{4 \pi^2}{q^2}\,n_V S_C({\mathbf q})\,
[1+2 X({\mathbf q})].
\label{pedecu}
\end{equation}
\\
In order to evaluate $\chi$ according to (\ref{limit}), we have to 
expand $C$ in powers of $q$. First, the above mentioned 
cancellation of $C(q=0)$ in (\ref{ddJdAdA}) with the (first) 
diamagnetic contribution is fulfilled, provided that the
effective charge of the Coulomb gas structure factor is chosen to be
\begin{equation}
q_V^2=2\pi J_{\parallel}\,(1+2 n_3)
\end{equation}
which is a reasonable generalization of the purely 2D result 
(\ref{charge}), taking into account the fact that the ``charges''
are now vortex lines elements correlated over a distance 
$\sim n_3d$ along the $z$-direction. The cancellation of the $q=0$ term 
in Eq.(\ref{ddJdAdA}) guarantees that the phase system of the 
superconductor has no stiffness above the critical temperature by 
making the limit (\ref{limit}) finite. Evaluating the latter
leads directly to the final result for the bulk susceptibility $\chi$ 
(per unit volume) which reads
\begin{equation}
\chi=-\frac{1}{(1+2n_3)d}\,\frac{k_{\!B}T}{\Phi_0^2}\,
[\frac{1}{n_V}+(\pi \gamma)^2 n_3(1+n_3)d^2].
\label{suscept}
\end{equation}
\\
Expression (\ref{suscept}) will be used in the following section for 
interpreting the experimental data (low field susceptibility and 
field dependence of the magnetization) obtained on underdoped 
YBCO.\cite{carretta}

\section{Analysis of experimental magnetization and susceptibility data}

In order to apply our theoretical results to underdoped YBCO, we use
the following (approximate) values for the lattice parameters :
$a=4$\AA\ and $d=12$\AA. Moreover, the molar quantities like those
reported in Ref.6 are obtained by multiplying 
expression (\ref{suscept}) by the molar volume $\Omega_M=
{\mathcal N}_{\!A} a^2d=115\,\text{cm}^3$ and by a reduction factor 
$\lambda$ which accounts for the fact that in the materials we are
interested in only a fraction of the unit cell volume actually 
carries the current densities and thus contributes to the 
diamagnetic response. 

\subsection{Low-field susceptibility}

\begin{figure}
\includegraphics[width=13cm]{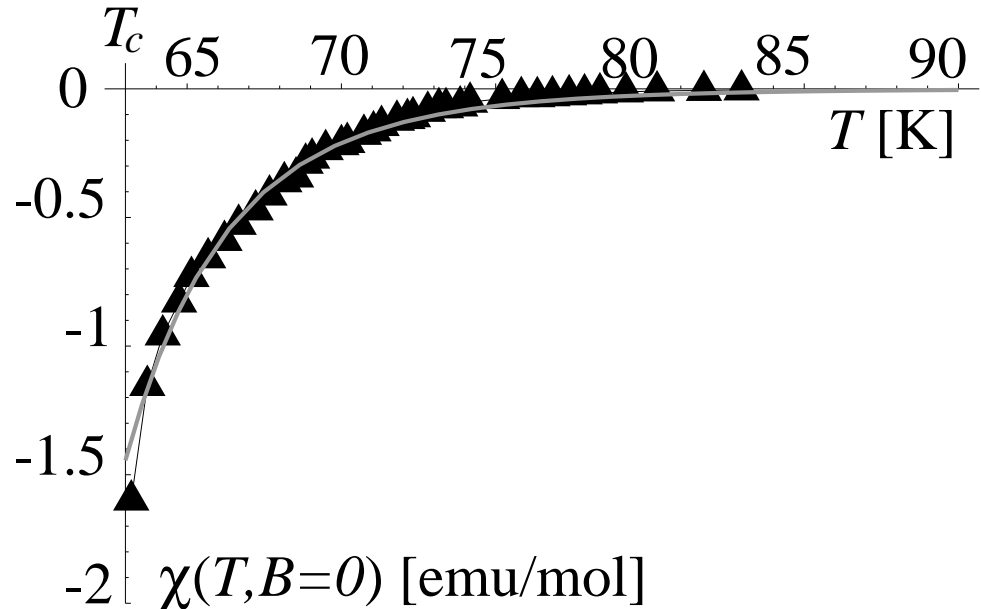}
\caption{Zero-field susceptibility $\chi(T,B=0)$ : the triangles 
($\blacktriangle$) are the experimental data from Ref.6 and 
the full line is the best fit of the theoretical expression 
(\ref{suscept}) assuming the activated behaviour (\ref{activ}).}
\label{fig2}
\end{figure}

The susceptibility data of Ref.6, for a very 
low applied field of 0.02T, cover the range from $T$ = 63K, which 
is just above the zero-field transition temperature, to 
$T$ = 110K. Above $T$ = 80K the diamagnetic susceptibility $\chi$
is essentially equal to zero, a background consisting of spin 
susceptibility and free-electron orbital diamagnetism having been 
subtracted. For zero applied field, our expression (\ref{suscept}) 
for $\chi$ contains several yet undetermined parameters: the areal 
density $n_V$ of vertical vortex line elements that are thermally excited,
the correlation length $n_3 d$ and the anisotropy $\gamma$. 
We mention that the subsequent analysis of the temperature and field 
dependent magnetization will show that the second term of (\ref{suscept}) 
is irrelevant at low fields. Therefore the value of $\gamma$ is
unimportant and the dominant contribution to $\chi(T,B\rightarrow 0)$
is given by in-plane correlations of the thermal vortex loop elements.
Fig.\ \ref{fig2}, in which the data of Ref.6 are 
reproduced with our expression (\ref{suscept}), shows that the observed 
zero-field diamagnetic susceptibility is almost perfectly fitted by
\begin{equation}
\chi(T,B=0)=C\exp(\frac{E_0}{k_{\!B}T})
\end{equation}
with $E_0/k_{\! B}T_c\approx 22$. This means that the quantity $n_V$
must be temperature dependent and obey a thermally activated behaviour
\begin{equation}
n_V = n_0\,\exp(-\frac{E_0}{k_{\!B}T})
\label{activ}
\end{equation}
with the factor $n_0\sim 10^{4}/a^2$. The number of vortex 
excitations in the XY-model is indeed known to show an activated 
$T$-dependence. In 2D this has recently been confirmed \cite{sengupta}
and various authors find the same result in 3D \cite{kohring,epiney,janke}
using numerical studies. In the first case the activation energy 
is roughly given by $E_0\sim$ 10 $k_{\!B} T_c$.\cite{sengupta}  
Simulations for the anisotropic 3D case yield values that are somewhat 
larger \cite{kohring} and that depend on the structure of the 
corresponding loop,\cite{epiney,janke,schmidt,carniero,tanner,nguyen,hu} 
getting larger the more planes are crossed. In a strongly 
anisotropic 3D XY-system, most of the thermal loops existing 
up to the critical temperature consist of elements parallel to
the lattice planes.\cite{chatto,janke,carniero,nguyen,hu} Loop
segments perpendicular to the planes only begin to be formed above
$T_c$. Thus the activated form attributed to the vortex line density
$n_V$ in (\ref{activ}) can be seen as a measure for the rapidly increasing 
number of loops crossing two planes or even more. Such loops contain 
at least two line segments perpendicular to the planes, the rest being
between two planes. This 3D structure may explain the 
relatively large value of the activation energy $E_0$ found above. 
The latter is also compatible with an explicit expression 
for the loop self-energy proposed in Ref.7 assuming that 
the total length of the loop segments oriented in $z$-direction 
is on the order of two interplanar distances, the rest of the loop 
being oriented parallel to the lattice planes. The value of $E_0$ 
further suggests that the quantity $n_3 d$ measuring the extension 
of the correlations of the vortex structure along the $z$-direction 
must be of the order of a few lattice distance $d$ (we will take
$n_3=2$ in the following). This provides an a posteriori 
justification for the assumptions we made when deriving 
Eq.(\ref{pedecu}). Finally we note that the fact that the
diamagnetic response is associated with loops containing line elements 
parallel to the $z$-direction directly follows from the magnetostatics:
the vertical loops elements are generated by currents flowing 
perpendicular to the external field $\mathbf B$ and thus 
responding the most sensitively to the latter.\\
We can now estimate roughly the value of the density of the 
vortices contributing to the diamagnetic response in the
temperature range of interest. Using the above values of $E_0$ 
and $n_0$ and taking into account a reduction factor of 0.15 
(see below), we get $n_V \sim 10^2$ $\mu$m$^{-2}$ for $T=65$K. 
This relatively small value will be discussed in the next section 
but we should not forget the fact that it concerns only the part of
the total number of vortices which contains elements perpendicular to 
the layers. We also remark that $\sqrt{n_V^{-1}}\gg a$, which fully 
justifies the continuous approach in the lattice planes used in 
the LD-action (\ref{XYaction}). 

\subsection{Field dependence of the magnetization}

\begin{figure}
\includegraphics[width=15cm]{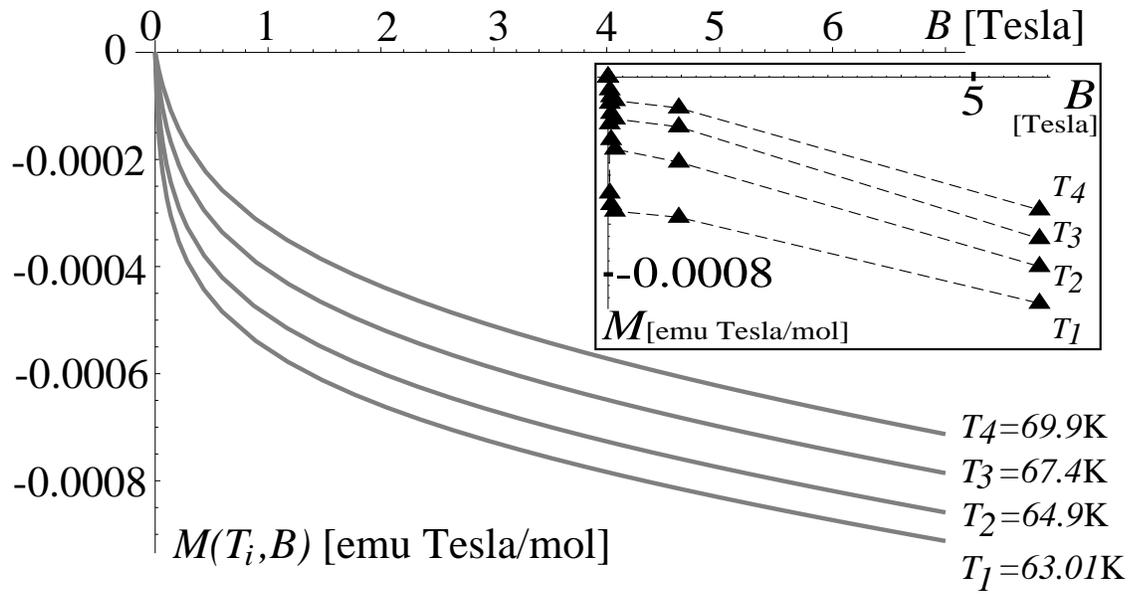}
\caption{Magnetization $M(T=\text{const},B)$ : the full lines
correspond to the theoretical expression (\ref{magnet}) with the
values of the parameters mentioned in the text and the triangles 
($\blacktriangle$) in the inset are the experimental data from Ref.6.}
\label{fig3}
\end{figure}

\begin{figure}
\includegraphics[width=15cm]{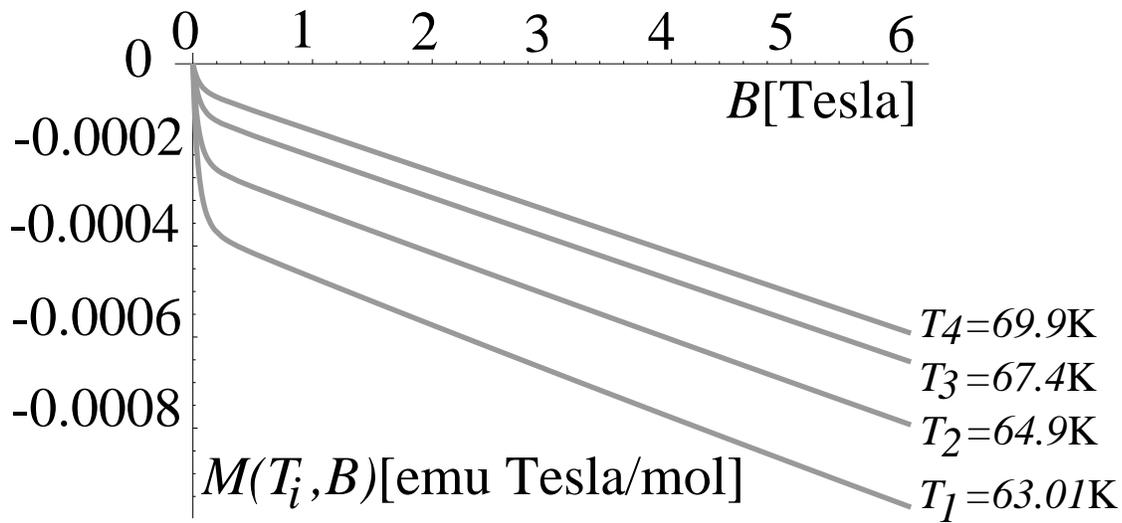}
\caption{Improved version of the magnetization $M(T=\text{const},B)$ :
the full lines correspond to the theoretical expression based on 
Eq.(\ref{improv}).}
\label{fig4}
\end{figure}

Here we have to deal with coexisting thermal loops and field induced
vortex lines.
\\
The magnetic field dependence in expression (\ref{suscept}) for 
$\chi$ is hidden in the density $n_V$ and, possibly, in $n_3 d$, 
the interplanar correlation length. The most simple approach to 
deal with this situation consists in splitting $n_V$ into a thermal 
and a field induced part as follows:
\begin{equation}
n_V=n_V^{th}+n_V^{f}
\equiv n_V^{th}[1+z(T,B)].
\end{equation}
This gives naturally rise to the dimensionless variable
\begin{equation}
z=\frac{n_V^{f}}{n_V^{th}}
=\frac{B}{\Phi_0\,n_V^{th}},
\label{scaling}
\end{equation}
quantifying the relative importance of the two types of vortex 
elements. Assuming that $n_3$ does not depend on 
$B$, one can integrate $\chi(T,B)$ in order to obtain the 
magnetization per unit volume:
\begin{equation}
M(T,B)=-\frac{1}{(1+2n_3)d}\,\frac{k_{\!B}T}{\Phi_0^2}
[\Phi_0\,\log {\boldsymbol (}1+z(T,B){\boldsymbol )}
+(\pi\gamma)^2 n_3 (1+n_3)d^2 B].
\label{magnet}
\end{equation}
The corresponding molar quantity is multiplied by the additional 
prefactor $\lambda \Omega_M$ already mentioned at the 
beginning of this section. $M$ depends on temperature through 
$n_V^{th}$ and, possibly, through $n_3$. The first contribution 
to (\ref{magnet}), given by $\log(1+z)$, has two different
limiting behaviours. For low fields $\log(1+z)\sim z$ and $M$ is
essentially given by $\chi(T,B=0)\,B$, whereas for high fields 
the first part of $M$ varies like $\log(z)$. The crossover 
between the two regimes takes place for $z\sim 1$ which describes the
situation where the quantities of thermally excited and field-induced 
vortex line elements are similar to one another. Thus the corresponding 
field value depends crucially on the thermal vortex element density 
$n_V^{th}$
showing up in the dimensionless variable $z$ defined in (\ref{scaling}). 
Using the values obtained in subsection III.A for the parameters $n_0$ 
and $E_0$ entering $n_V^{th}$, and choosing $\lambda$ such 
that the theoretical curves and the experimental data match 
reasonably, we find that $z\sim 1$ corresponds to a field $B_c\sim 0.025$T.
This is shown on Fig.\ref{fig3} where we note that the experimentally
observed behaviour shows indeed a crossover at fields of the order of 
0.05T. Therefore our theoretical crossover field $B_c$ is 
bit too small. However the two different regimes observed experimentally
can be clearly interpreted in the framework of our theory: they
describe the two situations where the diamagnetic response is essentially
due to solely one type of vortex line elements (thermal for small $B$ and
field induced for large $B$). This is already gratifying for a first approach.
For larger fields the term $\log(z)$ would yield a magnetization
that increases more slowly with $B$ than the measured data, which rather
show a linear $B$-dependence. This is reproduced by the second
part of (\ref{magnet}) which, in our procedure, arises from the
vortex correlations between different layers and becomes relevant 
for large values of $B$.\\
These considerations are illustrated on Fig.\ref{fig3} where the data from 
Ref.6 are shown together with the theoretical curves given by 
Eq.(\ref{magnet}). Although the quantitative agreement is less spectacular
than in the case of the zero-field susceptibility, it still allows to
extract the values of the ``magnetically active volume fraction'' $\lambda$
(from the small $B$ region) and of the anisotropy parameter $\gamma$ 
(from the large $B$ region). For the four temperatures of interest we find 
that the choice $\lambda\sim 0.15$ is the most satisfying. This is quite
reasonable for a layered compound such as underdoped YBCO where 
superconductivity occurs only in copper oxide planes which represent only
a small fraction of the unit cell. Concerning the anisotropy $\gamma$, we
find a value $\sim 2$. Its order of magnitude is correct since, by 
multiplying it by the lattice anisotropy $d/a\sim 3$, we obtain a value of 
6 for the effective anisotropy. The latter is somewhat lower than what is
observed in penetration depth measurements where it is found to be of the
order of $25$ at $T_c$.\cite{schneider} However, we recall that $\gamma$
entered our theory through the vortex line effective stiffness in 
Eq.(\ref{sergeo}). This rather rough description of the vortex loops 
interlayer correlations and the assumption we made by keeping $n_3$ 
independent of the temperature may possibly explain the fact that 
the observed value is a bit different from 25.\\
Using the values of $\lambda$ and $\gamma$ discussed above, it is 
now possible to show that the second term in our expressions 
(\ref{suscept}) and (\ref{magnet}) is irrelevant for the low-field 
behaviour, as we have anticipated when fitting the zero-field 
susceptibility. Finally we note that the picture of the vortex 
structure emerging from the above analysis which consists in loops and 
lines loosing their interplanar correlations over a very short distance 
$n_3d$ agrees very well with the usual descriptions of a vortex
system at temperatures lying above $T_c$ and above the vortex 
melting and a possible vortex decoupling 
line.\cite{chatto,blatter,ryu,ching,dodgson,qiu}
\\
However, as Fig.\ref{fig3} shows, the experimentally observed 
crossover is extremely sharp - in fact much sharper than what we can 
obtain using our expression (\ref{magnet}). This observation calls for a 
more refined treatment where other effects of the magnetic field than
the mere creation of vortex lines must be taken into account.
\\
It is in fact possible to understand, in the framework of our approach,
what makes the crossover so sharp. In terms of the susceptibility
$\chi(T,B)$ in Eq.(\ref{suscept}), the sharpness of the crossover
suggests that the first term must tend to zero very rapidly as $B$
increases. In this way, only the second (approximately constant) term
remains and then yields the linear behaviour of the magnetization 
$M(T,B)$ for $B>B_c$.
\\
This behaviour could be achieved by assuming a small magnetic field 
dependence of the activation energy $E_0$ and be motivated as follows.
In Ref.21 it is emphasized that, below $T_c$, the 
effective interaction between vortex loops is screened by the 
thermal defects of the Abrikosov vortex lattice. For small $B$, this 
effect is enhanced when the magnetic field is increased because the
density of  vortex lines is directly proportional to $B$ as in 
Eq.(\ref{scaling}). Above a crossover field $B^*$ the screening 
becomes weaker due to the fact that the finite stiffness 
($\propto B^2$) of the vortex lines inhibits the formation of further
defects. In the liquid phase above $T_c$ which we are interested in,
a similar qualitative behaviour can be expected, at least for low 
fields. A screened interaction between vortex loops then reduces 
the energetic cost of creating such an object. Thus the total number
of thermally excited vortex loops must increase with the magnetic 
field $B$. Among them a (small) percentage corresponds to those 
contributing to the diamagnetic response as discussed in the
previous section. They will also follow the above behaviour so that
we may reasonably assume that their effective activation energy $E_0$ 
introduced in Eq.(\ref{activ}) must decrease slightly when the
magnetic field $B$ increases. To lowest order, we have
\begin{equation}
E_0\rightarrow E_0(B)=\left\{
\begin{array}{ll}
E_0(1-\alpha B),& B\ll B_c\\
E_1<E_0,        & B\gg B_c.
\end{array}
\right.
\label{improv}
\end{equation}
The large $B$ behaviour is not important since it affects only the
first term of the susceptibility when it is already very small. To 
illustrate this idea, we take $\alpha=1$ such that the value 
of the activation energy saturates at a value $E_1$ 20 percents lower 
than $E_0$ around $B=0.5T$. After having performed a numerical integration
of the susceptibility $\chi(T,B)$ over $B$, we obtain the curves 
shown in Fig.\ref{fig4}. The crossover between the two field regimes 
is sharper than in the previous case and the corresponding value 
of the magnetic field $B_c$ is higher.\\
We emphasize however that the above strategy was only intended to
show how the developed formalism could be modified in order to 
improve the interpretation of the experimental results from 
Ref.6. A proper justification of the above
assumption (\ref{improv}) requires concepts and methods which
are beyond the scope of this work.

\section{Summary and conclusions}

We have derived explicit expressions for the field and temperature
dependent diamagnetic susceptibility $\chi(T,B)$ and magnetization
$M(T,B)$ of an anisotropic superconductor above its critical 
temperature. The superconducting fluctuations above $T_c$ are 
treated in the framework of the Lawrence-Doniach model including a
magnetic field $\mathbf B$ perpendicular to the lattice planes.
In order to describe specifically the precursor effects above $T_c$,
we have used the London approximation, assuming thereby that the 
relevant fluctuations are given by the phase of the order parameter 
only. Such an approach should apply to the underdoped regime of 
cuprates. In this context the current-current correlation function 
that determines $\chi$ and $M$ is expressed by the structure factor
of the relevant phase excitations - the thermally excited vortex 
loops and the field induced vortex lines. Our expressions for $\chi$
and $M$ contain still several undetermined parameters such as the areal
density of vortex line elements within a given layer, an appropriate 
length describing vortex line correlations between different layers,
the anisotropy and a factor smaller than one that gives the 
fraction of the volume of the superconductor that is ``active'' 
in contributing to the fluctuation induced diamagnetism. These 
quantities, which must be compatible to another, are 
estimated by comparing $\chi(T,B=0)$ and $M(T,B)$ to experimental 
data obtained on underdoped YBCO \cite{carretta}and allow then to deduce 
an accurate picture of the phase system of such materials.\\
Our findings can be summarized as follows:
\\
\textit{(i)} There have been attempts to understand the same data based
on Gaussian fluctuations of the anisotropic LD-model.\cite{carretta} The 
corresponding fit for optimally doped YBCO, where the precursor
region is much narrower, has worked out successfully. However, the 
discrepancies between experiment and the corresponding theoretical 
curves for the underdoped compound are very large, both in the 
temperature and magnetic field dependences. Our approach, based on 
topological phase excitations in a non critical regime, is much more 
satisfactory.
\\
\textit{(ii)}
The experimental low-field susceptibility has an activated temperature
behaviour. This observation indeed points to excitations which have a
finite creation energy (the vortex loops), rather than some wave-like
fluctuations of the pairing field (the Gaussian modes). The value of 
the activation energy is roughly 20 times the critical temperature, 
what shows that the relevant vortex loops extend over several 
layers. The density of those loops is relatively small, indicating 
that, close to the critical temperature, the phase system consists 
predominantly of vortex loops running essentially parallel to the 
planes, and the line elements in $z$-direction which contribute 
to $\chi$ only develop significantly above the critical temperature.
\\
\textit{(iii)}
The interlayer current correlations reflecting the geometrical
structure of vortex loops in the $z$-direction are taken into
account in terms of an effective correlation length $\xi_3=n_3 d$. The
value of the latter has been determined by the vortex loop activation
energy. In the temperature range of the measurements it is small, 
typically 2 or 3 interlayer distances. This feature can be also
characterized by the stiffness of the vortex line which is found
to be approximately inversely proportional to the anisotropy
parameter. These observations indicate that the observed phase 
structure corresponds very well to the picture of weakly coupled 
layers of two-dimensional ``pancakes'' vortices.
\\
\textit{(iv)}
In the calculation of the molar susceptibility we have multiplied the
molar volume by an overall prefactor called ``effective active volume
fraction'' taking into account only those parts of the unit cell that 
contribute to superconducting fluctuations. This quantity was found to
be of the order of 10-20 percents.
\\
\textit{(v)}
Both the activated behaviour of $\chi$ and the small value of the above 
correlation length $\xi_3=n_3 d$ point to the fact that the 
measurements of Ref.6 cover a ``precritical'' regime. Indeed, although 
the relevant physics is governed by phase fluctuations according to our
assumptions, no divergence of the 3D XY-type has been observed. The
latter is nevertheless expected for the data taken the closest to $T_c$,
as it has been identified in Ref.8 for layered superconductors with a
reasonable value of the anisotropy. In fact one observes that some
deviations from the activated behaviour appear for the temperatures the
closest to $T_c$. This may indicate a crossover to a critical regime in
which the $T$-dependence of $\chi$ is no more simply governed by the
\emph{total} vortex density, but by a diverging correlation length $\xi$
associated with the \emph{free} vortex density which should be used
instead of $n_V$ in the structure factor (\ref{coulomb}) and in
$\chi$ [Eq.(\ref{suscept})], as it was done in the 2D case for
calculating the zero field susceptibility.\cite{halperin} Moreover,
in this case the dimensionless quantity $z$ defined in
Eq.(\ref{scaling}) becomes the scaling variable $B\xi^2/\Phi_0$ which
has been used extensively to study critical properties under magnetic
field.\cite{schneider}
\\
\textit{(vi)}
The experimental data show an extremely sharp crossover in the field
dependence of the magnetization. Within our approach this is
attributed to a subtle interplay between the two types of vortices 
suggesting that the presence of field-induced vortex lines is 
favoring the creation of thermal vortex loops containing at least
two segments along the $z$-direction. However, even our most refined 
theoretical curves are still more smooth than the data. It is 
therefore not excluded that the observed behaviour may have yet 
another origin than superconducting fluctuations, for instanced 
sample inhomogeneities, as it has been suggested and observed in
other samples by the authors of the experimental work \cite{carretta}
that we have analysed.
\\
We thank X. Zotos, A.A. Varlamov, S.R. Shenoy and T. Schneider for interesting
discussions. This work has been supported by the Swiss National
Science Foundation (projects  20-49586.96 and 2000-053697.98/1).
%
%

%
%
\newpage
\appendix
\section*{}
Here we show that the zero wave vector value of the positional 
structure factor (\ref{strucfac}) compensates the singular term
of the current-current correlation function (\ref{curcurcor}).
\\
Starting from Eq.(\ref{strucfac}), we find
\begin{equation}
\begin{split}
S({\mathbf q}=0)&=\sum_{n,s,n'\!,s'} 
t(s,n)\,t(s',n') \\
&=\sum_{n,n'}(N_+ - N_-)^2
\end{split}
\label{A1}
\end{equation}
where $N_+$ [$N_-$] is the number of vortex line elements with
$t(s,n)=+1$ [$t(s,n)=-1$] in layer $n$. They correspond exactly to
the 2D concepts of vortex and antivortex which are often interpreted as
positive and negative charges of a 2D Coulomb gas.
\\
The external flux generates $N_F$ ``charges'' according to
\begin{equation}
N_F=\frac{L^2 B}{\Phi_0}.
\end{equation}
Due to the incompressibility of the magnetic field  
(${\boldsymbol\nabla}\!\cdot\!{\mathbf B}=0$), this number 
is the same for all layers.\\
The neutrality condition requires that, in every layer $n$, the external 
flux is exactly compensated by the net charge $N_+ - N_-$ of the 
vortex-antivortex system:
\begin{equation}
N_F=N_+ - N_-,\,\,\forall n.
\end{equation}
Therefore (\ref{A1}) may be written as
\begin{equation}
S({\mathbf q}=0)=N^2\big(\frac{L^2 B}{\Phi_0}\big)^2
\end{equation}
which is exactly the singular term of the current-current correlation
function $C({\mathbf q})$.
%
%
\newpage
\newpage

\end{document}